\documentclass[10pt,twoside,twocolumn,english,aps,amssymb,amsmath,amsfonts,superscriptaddress,showpacs,prb]{revtex4}
\usepackage[T1]{fontenc}
\usepackage[latin9]{inputenc}
\usepackage[active]{srcltx}
\usepackage{babel}
\usepackage{array}
\usepackage{textcomp}
\usepackage{multirow}
\usepackage{amstext}
\usepackage{amssymb}
\usepackage{graphicx}
\usepackage{esint}
\usepackage[usenames,dvipsnames]{xcolor}
\usepackage{mathrsfs,ulem,epstopdf}
\usepackage{bm}
\usepackage{wasysym}
\usepackage{color}
\usepackage[unicode,pdfusetitle, 
 bookmarks=false,bookmarksnumbered=false,bookmarksopen=false,
 breaklinks=false,pdfborder={0 0 1},backref=false, colorlinks, 
 citecolor=red,linkcolor=blue,urlcolor=NavyBlue]{hyperref}
\usepackage{breakurl}

\makeatletter

\DeclareRobustCommand{\greektext}{%
  \fontencoding{LGR}\selectfont\def\encodingdefault{LGR}}
\DeclareRobustCommand{\textgreek}[1]{\leavevmode{\greektext #1}}
\DeclareFontEncoding{LGR}{}{}
\DeclareTextSymbol{\~}{LGR}{126}
\newcommand{\lyxmathsym}[1]{\ifmmode\begingroup\def\b@ld{bold}
  \text{\ifx\math@version\b@ld\bfseries\fi#1}\endgroup\else#1\fi}

\providecommand{\tabularnewline}{\\}

\@ifundefined{textcolor}{}
{%
 \definecolor{BLACK}{gray}{0}
 \definecolor{WHITE}{gray}{1}
 \definecolor{RED}{rgb}{1,0,0}
 \definecolor{GREEN}{rgb}{0,1,0}
 \definecolor{BLUE}{rgb}{0,0,1}
 \definecolor{CYAN}{cmyk}{1,0,0,0}
 \definecolor{MAGENTA}{cmyk}{0,1,0,0}
 \definecolor{YELLOW}{cmyk}{0,0,1,0}
}

\makeatother

\begin{document}

\title{NMR, magnetization, and heat capacity studies of the uniform\\ 
spin-$\frac{1}{2} $ chain compound Bi$_{6}$V$_{3}$O$_{16}$}

\author{T. Chakrabarty}
\email{tanmoy.chakrabarty@kbfi.ee}
\affiliation{National Institute of Chemical Physics and Biophysics, 12618 Tallinn,
Estonia}
\affiliation{Tata Institute of Fundamental Research, Homi Bhabha Road, Colaba,
 400005 Mumbai, India}
\author{I. Heinmaa}
\email{ivo.heinmaa@kbfi.ee}
\affiliation{National Institute of Chemical Physics and Biophysics, 12618 Tallinn,
Estonia}
\author{V. Yu. Verchenko}
\email{valeriy.verchenko@gmail.com}
\affiliation{National Institute of Chemical Physics and Biophysics, 12618 Tallinn,
Estonia}
\affiliation{Department of Chemistry, Lomonosov Moscow State University, 119991
Moscow, Russia}
\author{P. L. Paulose}
\email{paulose@tifr.res.in}
\affiliation{Tata Institute of Fundamental Research, Homi Bhabha Road, Colaba,
 400005 Mumbai, India}
\author{R. Stern}
\email{raivo.stern@kbfi.ee}
\affiliation{National Institute of Chemical Physics and Biophysics, 12618 Tallinn,
Estonia}
\received {29 January 2019; Revised 1 August 2019}
\begin{abstract}
We report the local (NMR) and bulk (magnetization and heat capacity)
properties of the vanadium based spin-$\frac{1}{2} $ uniform spin chain compound Bi$_{6}$V$_{3}$O$_{16}$
(Bi$_{4}$V$_{2}$O$_{10.66}$). In the low-temperature $\alpha$ phase, the
magnetic ions (V$^{4+}$) are arranged in one-dimensional chains.
The magnetic susceptibility shows a broad maximum around $50$~K 
signifying a short-range magnetic order. 
Heat capacity measurements also reveal low-dimensional magnetism. The $^{51}$V magic
angle spinning nuclear magnetic resonance measurements
clearly show that the magnetic V$^{4+}$ and non-magnetic V$^{5+}$
species are located on different crystallographic sites with no mixed occupation.
The spin susceptibility calculated from
the shift of the $^{51}$V NMR spectra reproduces the behavior
observed in magnetic susceptibility and agrees well with the spin-$\frac{1}{2} $
uniform spin chain model with $J = 113(5)$~K. 
\end{abstract}

\pacs{76.60.-k, 75.47.Lx, 75.50.Ee}

\maketitle

\section{introduction}

Magnetism in one-dimensional (1D) Heisenberg antiferromagnetic
(AFM) spin systems has remained an area of wide interest in condensed-matter physics since 1970s.~\cite{Lieb1961,Affleck1989}
This is mainly due to the rich physics such spin chains exhibit. Furthermore,
these systems are tractable from both theoretical~\cite{Eggert1994,KuJohnston2000}
and computational~\cite{BF1964,Bonner1979,Alternating chain expression paper_Johnston}
starting points. The prime interest here is the innate strong quantum
fluctuations which lead to the suppression
of magnetic long range ordering.~\cite{Mermin Wagner prevention of LRO in 1D}
The nature of the ground state in these systems depends on the value
of spin $S$ and relative strength and sign (AFM or ferromagnetic)
of their coupling $J$. The generic magnetic Hamiltonian describing
an $S=\frac{1}{2}$ Heisenberg chain can be written as $H~=~J\sum_{i}S_{i}S_{i+1}$,
where $J$ is the intrachain coupling constant between the nearest-neighbor
spins. A uniform half-integer spin chain with AFM interactions
exhibits a gapless ground state.~\cite{Endoh1974,Nagler1991,Tennant1993,Lake2005,SrCu2O3,Ronnow2013,Ba2Cu(PO4)2-NMR-T1,NOCu(NO3)3-1,NOCu(NO3)3-2,NOCu(NO3)3-3,Tran2019}
In the case of an AFM alternating Heisenberg chain, the AFM exchange
constants ($J_{1}$ and $J_{2}$) between the two nearest-neighbor
spins are unequal ($J_{1}$$\neq$ $J_{2}$; $J_{1}$, $J_{2}$$>0$),
and they alternate from bond to bond along the chain with an alteration
parameter $\alpha=J_{2}/J_{1}$. 
In the case of an alternating chain, interactions for half-integer spins result in the 
opening of a gap via either frustration due to the next-nearest-neighbor AFM 
exchange or dimerization due to the alternating coupling to nearest neighbors 
along the chain.~\cite{spin gap in alternating spin chain,Bonner1983,
Xu2000,VOPO_Johnston,VOPO_Barnes,VOPO_Garrett,(VO)2P2O7-NR_T1_spin gap,CaCuGe2O6,AgVOAsO4 NMR,Lebernegg2017}
Many $S=\frac{1}{2}$ chain systems have already been reported, and their
characteristics have been interpreted on the basis of such a model.
Some of the most renowned examples in this category are CuCl$_{2}\cdot$2N(C$_{5}$D$_{5}$),~\cite{Endoh1974}
KCuF$_{3}$,~\cite{Nagler1991,Tennant1993,Lake2005} Sr$_{2}$CuO$_{3}$,\cite{SrCu2O3} CuSO$_{4}\cdot$5H$_{2}$O,~\cite{Ronnow2013}, Ba$_{2}$Cu(PO$_{4}$)$_{2}$,~\cite{Ba2Cu(PO4)2-NMR-T1} and, recently, (NO)[Cu(NO$_3$)$_3$]~\cite{NOCu(NO3)3-1,NOCu(NO3)3-2,NOCu(NO3)3-3} and Cs$_4$CuSb$_2$Cl$_{12}$~\cite{Tran2019} for uniform and
Cu(NO$_{3}$)$_{2}\cdot$2.5H$_{2}$O,~\cite{Bonner1983,Xu2000}
(VO)$_{2}$P$_{2}$O$_{7}$,\cite{VOPO_Johnston,VOPO_Barnes,VOPO_Garrett,(VO)2P2O7-NR_T1_spin gap}
CaCuGe$_{2}$O$_{6}$,\cite{CaCuGe2O6} 
AgVOAsO$_{4}$\cite{AgVOAsO4 NMR}, and Cu$_3$(MoO$_4$)(OH)$_4$\cite{Lebernegg2017} 
for alternating $S=\frac{1}{2}$ Heisenberg spin chains.

\begin{figure}
\begin{centering}
\includegraphics[scale=0.056]{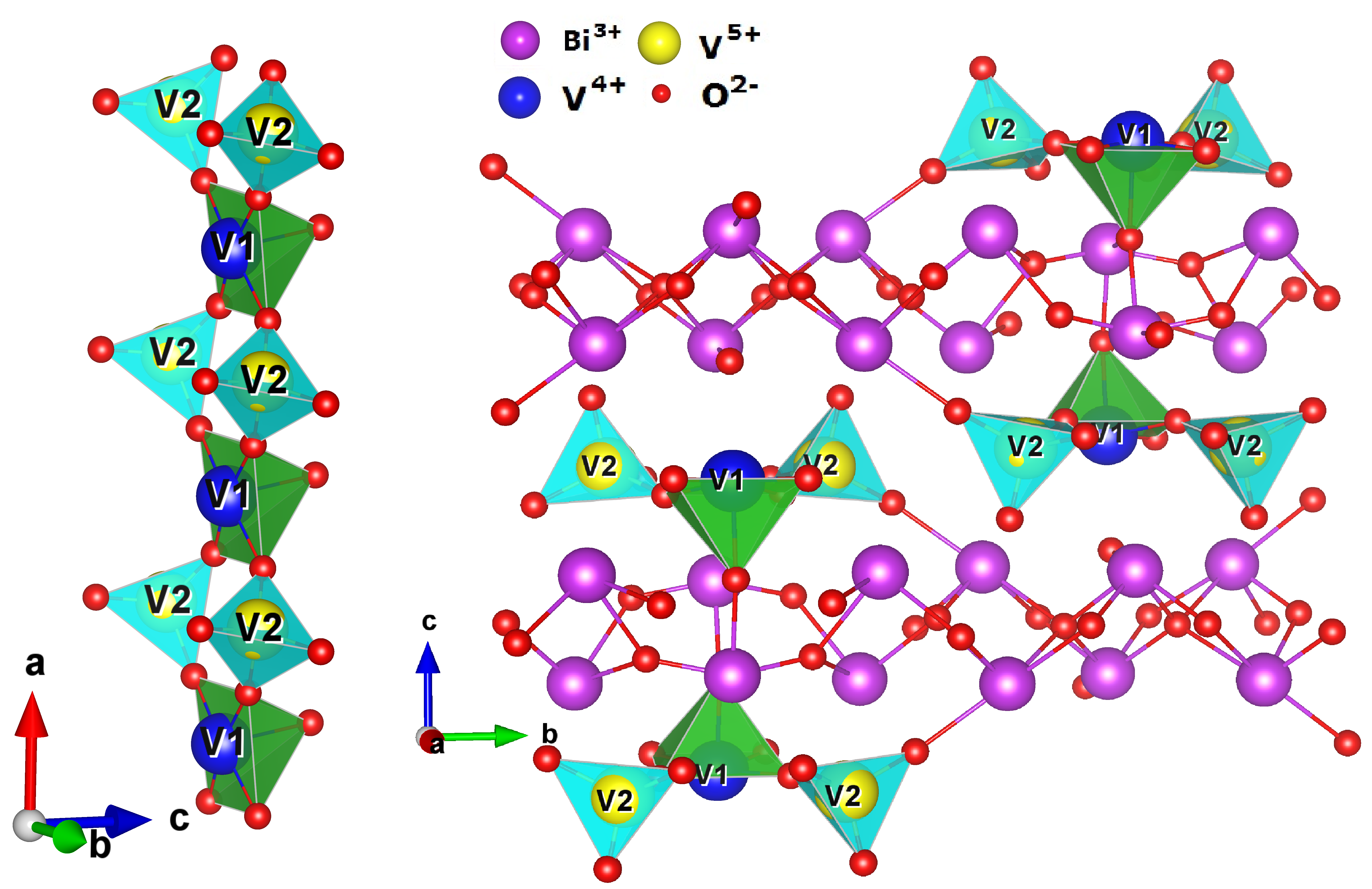} 
\par\end{centering}

\caption{\label{fig:1 uniform chain structure}Left: Possible interaction path
between the magnetic V$^{4+}$ ions in Bi$_{6}$V$_{3}$O$_{16}$
mediated via nonmagnetic V$^{5+}$ ions and oxygen. Right: Top view of
the chains separated by Bi-O layers. }
\end{figure}

Our motivation is to explore new low-dimensional magnetic oxides with
the intention of unraveling novel magnetic properties. In this paper
we report the bulk and local (NMR) studies of the vanadium-based $S=\frac{1}{2}$
uniform spin chain compound Bi$_{6}$V$_{3}$O$_{16}$ (often described also as
Bi$_{4}$V$_{2}$O$_{10.66}$) at low temperatures.
This system is a member of the well-known pseudo binary oxide systems
Bi$_{2}$O$_{3}$-V$_{2}$O$_{5}$, which received significant interest
because of their different structural varieties and rich functional
properties,~\cite{Reference for Bi2O3-V2O5,First report on Bi4V2O10}
which led also to a very efficient bismuth-metal-vanadia (BiMeVOX) family of anionic conductors.~\cite{BiMeVOX,SOEC-review}
These systems belong to or are derived from the Aurivillius family.~\cite{Aurivillius_1,Aurivillius_2}
They exhibit three polymorphs, $\alpha$, $\beta$, and $\gamma$, each
associated with a different temperature range, where the $\alpha$ phase
is the low-temperature one. One of these Aurivillius vanadates,
Bi$_{4}$V$_{2}$O$_{10}$ which contains all the vanadium ions in
the V$^{4+}$ oxidation state, was studied thoroughly via crystal
structure, electron diffraction, and thermodynamic properties about
two decades ago.\cite{Bi4V2O10_structure,Bi4V2O10_chain structure}
In the $\alpha$ phase of Bi$_{4}$V$_{2}$O$_{10}$, the magnetic
V$^{4+}$ ions are arranged in a 1D chain along the $a$ direction
of the unit cell (see Fig.~\ref{fig:1 uniform chain structure}; 
\href{http://jp-minerals.org/vesta/en/}{\textsc{vesta}}
software~\cite{vesta} was used for crystal structure visualization).
It was also proposed by Satto \textit{et al.}\cite{Bi4V2O10_chain structure}
that $\alpha-$Bi$_{4}$V$_{2}$O$_{10}$ transforms to $\alpha-$Bi$_{6}$V$_{3}$O$_{16}$
after oxidation upon exposure to air at room temperature. The orientation
of the magnetic V$^{4+}$ ions in 1D chains remains intact in the
crystal structure of the $\alpha$ phase of Bi$_{6}$V$_{3}$O$_{16}$,
which is best described by V$_{3}$O$_{16}^{6-}$ ribbons running along
the $a$ axis and containing units built up from a pyramid (V$^{4+}$)
and two tetrahedra (V$^{5+}$).~\cite{Bi4V2O10.66_crystal structure,Bi4V2O11 to Bi4V2O10.66}
This is one of the few systems found so far in nature where the extended superexchange 
interaction takes place by the overlap of $d_{xy}$ (via the oxygen $p$ ) orbitals of $d_1$ 
electrons ($t_{2g}$) of V$^{4+}$ ions rather than the $d_{x2-y2}$ orbitals of $e_g$ electrons 
in Cu-based systems, with an exchange coupling ($J/k_B$) as high as $\sim$~100~K.
Very recently, magnetic properties and charge ordering were reported
for another related compound, Bi$_{3.6}$V$_{2}$O$_{10}$,\cite{Ref.19_magnetic measurement-BiV2O5_2013,BiV2O5_Oct 2015_paper}
which also belongs to the aforementioned Aurivillius family. However,
a detailed investigation of the bulk and local properties with a proper
theoretical model has not been carried out for any of these magnetic
Aurivillius vanadates. This leads to the the primary sources of motivation 
for our present work.


\section{Sample preparation, crystal structure, and experimental details}

Bi$_{6}$V$_{3}$O$_{16}$ is an orthorhombic system and crystallizes in
the  $Pnma$ space group.~\cite{Bi4V2O10.66_crystal structure} The
low-temperature phase of Bi$_{6}$V$_{3}$O$_{16}$ was synthesized
by mixing stoichiometric amounts of Bi$_{2}$O$_{3}$ (Alfa Aesar, $99.99\%$)
and VO$_{2}$. VO$_{2}$ was prepared through the reaction of an equimolar mixture
of V$_{2}$O$_{5}$ and V$_{2}$O$_{3}$ at $680\lyxmathsym{\textdegree}$
C for $18$ h under vacuum. V$_{2}$O$_{3}$ was obtained by reducing
V$_{2}$O$_{5}$ ($99.99\%$, Aldrich) under hydrogen flow at $800$$\lyxmathsym{\textdegree}$
C. The mixture of Bi$_{2}$O$_{3}$ and VO$_{2}$ was then pelletized
and placed in a quartz ampule sealed in vacuum ($<10^{-5}$mbar). 
The ampule was annealed at $550$\textdegree{} C for $48$ h. This process was repeated
three times with intermediate grinding and mixing. The last round
of heating was performed at $620$\textdegree{} C for $36$ h.
After a few weeks of exposure in open air, Bi$_{4}$V$_{2}$O$_{10}$ self-oxidized 
into Bi$_{6}$V$_{3}$O$_{16}$, and its color changed from black to dark brown. 
A similar transformation was observed previously by Satto \textit{et al.}\cite{Bi4V2O10_chain structure}

\begin{figure}
\begin{centering}
\includegraphics[scale=0.5]{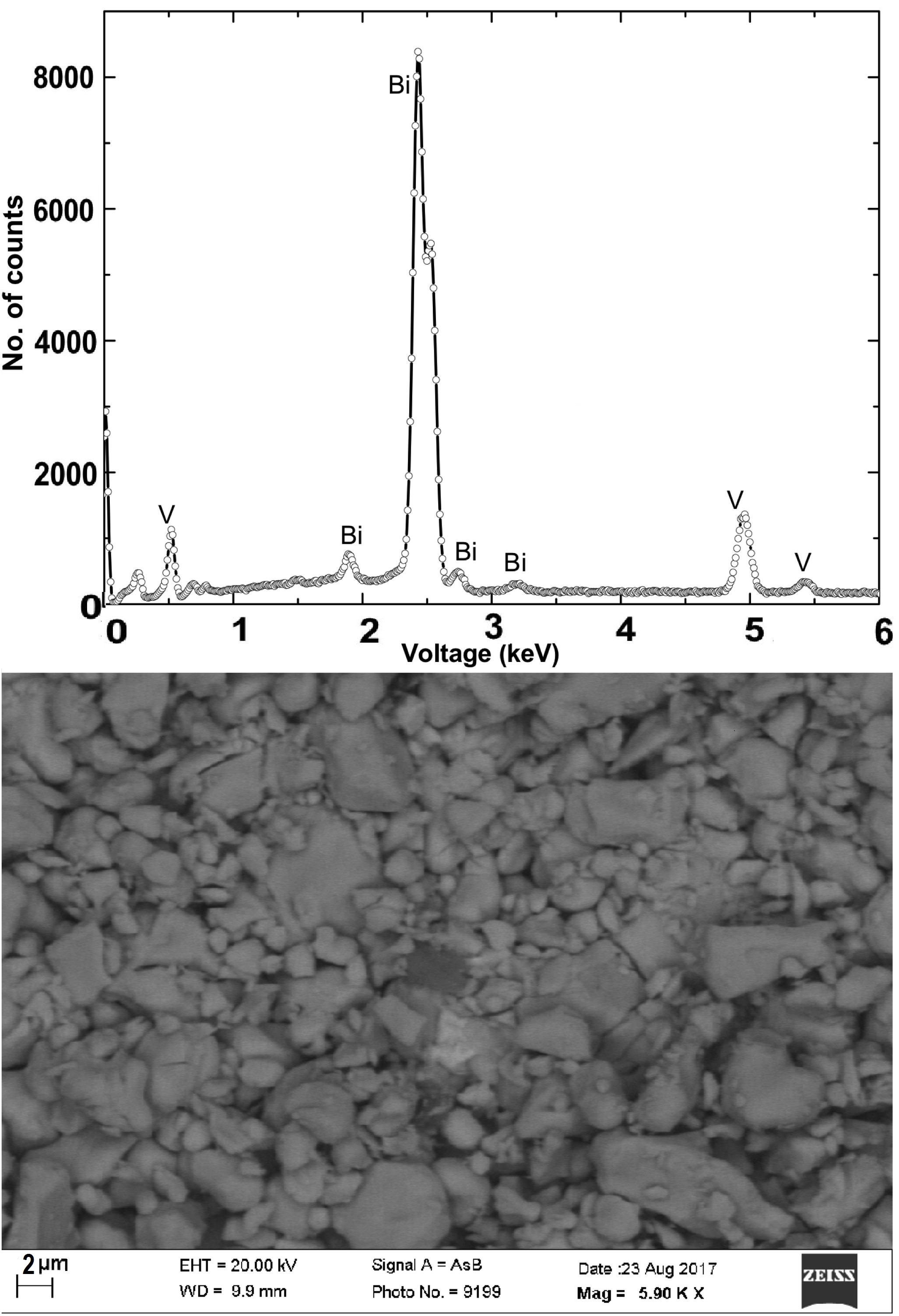} 
\par\end{centering}

\caption{\label{fig2:EDX&SEM_ Fig2}EDX spectrum (top) and SEM image (bottom)
of polycrystalline Bi$_{6}$V$_{3}$O$_{16}$.}
\end{figure}

Energy dispersive x-ray (EDX) microanalysis show an elemental ratio
of bismuth and vanadium of Bi:V$\simeq2:1$ (see
Fig.~\ref{fig2:EDX&SEM_ Fig2}).

\begin{figure}
\begin{centering}
\includegraphics[scale=0.11]{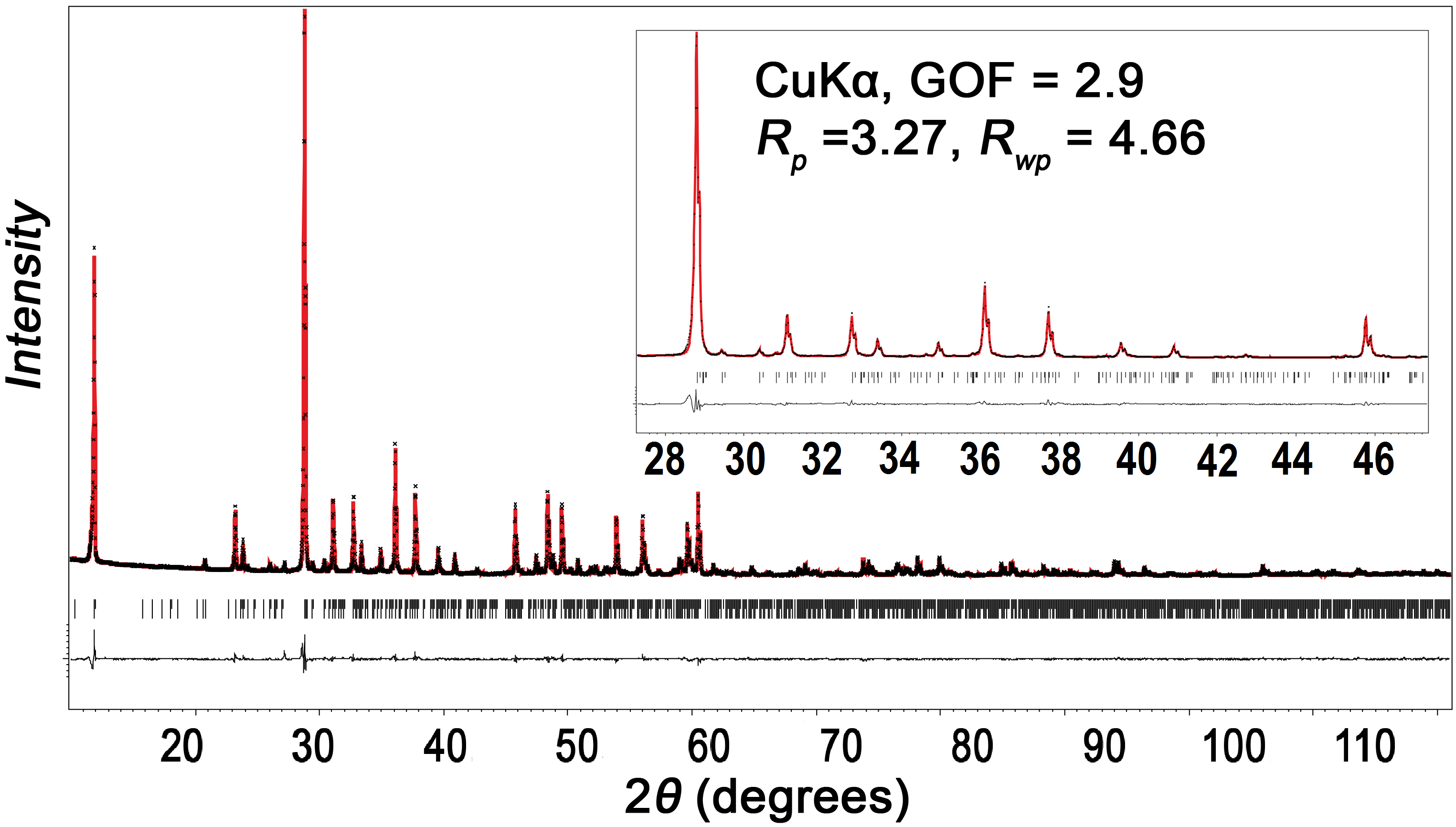} 
\par\end{centering}

\caption{\label{fig3:xrda+Alfa_refinement}
Experimental (black points) and calculated (red line) powder XRD patterns of Bi$_{6}$V$_{3}$O$_{16}$. 
Positions of peaks are given by black ticks, and the difference plot is shown
by the black line in the bottom part. }
\end{figure}

X-ray diffraction (XRD) patterns were collected using
a PANalytical X'Pert$^3$ Powder x-ray diffractometer (Cu $K \alpha$
radiation, $\lambda=1.54182$~\AA).
The 
Rietveld refinement against the XRD data was carried out using the \textsc{jana}2006
software\cite{JANA2006}(see Fig.~\ref{fig3:xrda+Alfa_refinement}).
The XRD pattern shows a single phase of Bi$_{6}$V$_{3}$O$_{16}$ (space
group: $Pnma~(62)$,\ 
$a=5.47124(3)$~\AA,\
$b=17.25633(8)$~\AA, \
$c=14.92409(6)$~~\AA). 
The ratio of $c/(b/3)\simeq2.6$ obtained from the refinement is consistent
with the previous study carried out on single crystals of the $\alpha$ phase 
of Bi$_{6}$V$_{3}$O$_{16}$.~\cite{Bi4V2O11 to Bi4V2O10.66}
The refined atomic coordinates of Bi$_{6}$V$_{3}$O$_{16}$ are given
in Table \ref{atomic positions_alfa}.

\begin{table}
\begin{centering}
\begin{tabular}{|c|c|c|c|c|}
\hline 
Atoms  & \multicolumn{1}{c}{} & \multicolumn{1}{c}{Coordinates} & \multicolumn{1}{c|}{}  & TDP \tabularnewline
\hline 
 & $x/a$  & $y/b$  & $z/c$  & $B_{iso}$ ($\textrm{\ensuremath{\mathring{A}}}^{2}$)\tabularnewline
\hline 
Bi1($4c$)  & $0.2004(7)$  & $0.250$  & $0.31668(17)$  & $2.13(7)$\tabularnewline
\hline 
Bi2($4c$)  & $0.2519(7)$  & $0.250$  & $0.66949(17)$  & $2.08(8)$\tabularnewline
\hline 
Bi3($8d$)  & $0.2368(4)$  & $0.41275(14)$  & $0.83414(13)$  & $1.98(6)$\tabularnewline
\hline 
Bi4($8d$)  & $0.2835(5)$  & $0.42850(12)$  & $0.16763(14)$  & $2.25(6)$\tabularnewline
\hline 
V1($4c$)  & $0.246(3)$  & $0.250$  & $0.0331(7)$  & $4.1(3)$\tabularnewline
\hline 
V2($8d$)  & $0.265(2)$  & $0.3864(4)$  & $0.4874(6)$  & $3.8(2)$\tabularnewline
\hline 
O1($4c$)  & $-0.030(4)$  & $-0.0171(15)$  & $0.2391(15)$  & $0.5(2)$\tabularnewline
\hline 
O2($8d$)  & $0.127(4)$  & $0.3523(13)$  & $0.2921(11)$  & $0.5(2)$\tabularnewline
\hline 
O3($8d$)  & $0.010(6)$  & $0.323(2)$  & $0.7464(15)$  & $0.5(2)$\tabularnewline
\hline 
O4($8d$)  & $-0.018(6)$  & $0.3226(17)$  & $0.0536(10)$  & $0.5(2)$\tabularnewline
\hline 
O5($4a$)  & $-0.015(6)$  & $0.331(2)$  & $0.4704(9)$  & $0.5(2)$\tabularnewline
\hline 
O6($8d$)  & $0.287(4)$  & $0.4056(14)$  & $0.5912(10)$  & $0.5(2)$\tabularnewline
\hline 
O7($8d$)  & $0.207(5)$  & $0.4749(11)$  & $0.4325(10)$  & $0.5(2)$\tabularnewline
\hline 
O8($8d$)  & $0.316(6)$  & $0.250$  & $0.1752(16)$  & $0.5(2)$\tabularnewline
\hline 
O9($8d$)  & $0.178(6)$  & $0.250$  & $0.8582(15)$  & $0.5(2)$\tabularnewline
\hline 
\end{tabular}
\par\end{centering}

\centering{}\caption{\label{atomic positions_alfa} Atomic coordinates and  thermal 
displacement parameters in the crystal structure of Bi$_{6}$V$_{3}$O$_{16}$.}
\end{table}

\begin{figure}
\begin{centering}
\includegraphics[scale=0.48]{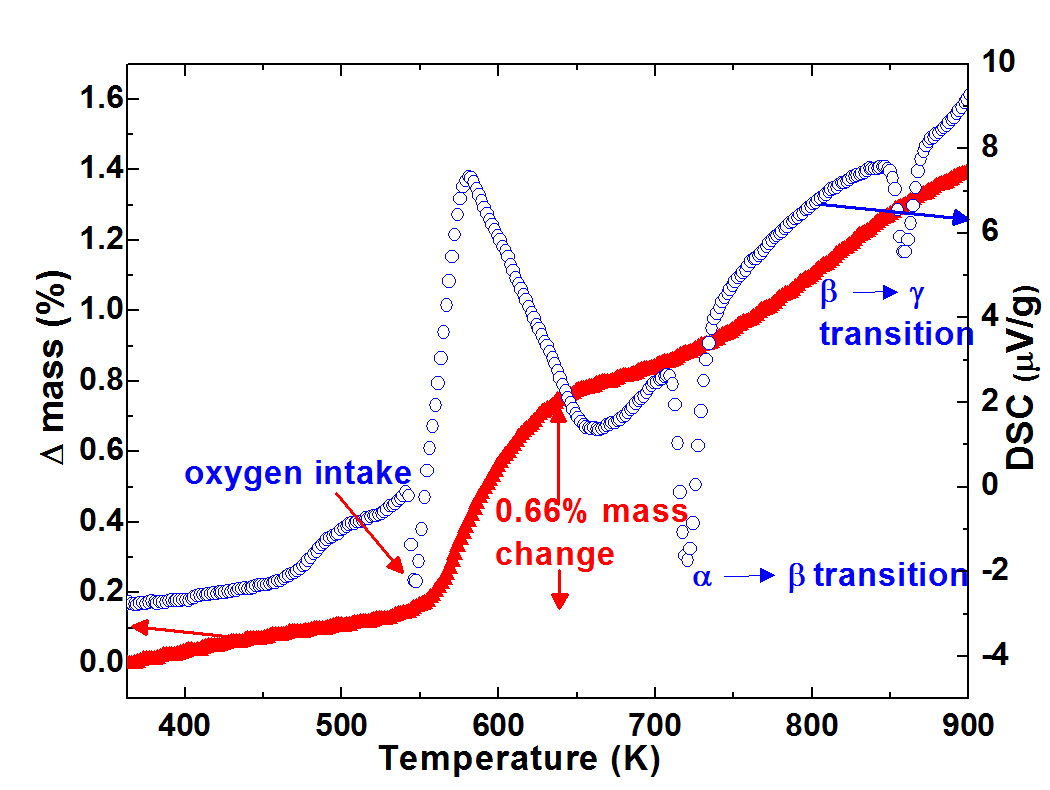} 
\par\end{centering}

\caption{\label{fig:TGA_heating}Temperature dependence of the mass change
and the heat flow during heating of Bi$_{6}$V$_{3}$O$_{16}$ in
the TGA experiment.}
\end{figure}

The thermogravimetric analysis (TGA) was carried out to heat
Bi$_{6}$V$_{3}$O$_{16}$ and to oxidize it while observing the weight change 
(see Fig.~\ref{fig:TGA_heating}). After the full oxidation, Bi$_{6}$V$_{3}$O$_{16}$ 
 (Bi$_{4}$V$_{2}$O$_{10.66}$) should transform into Bi$_{4}$V$_{2}$O$_{11}$
in which all vanadium ions are in the V$^{5+}$ oxidation state. In our experiments,
we observed three major temperature effects. The first one at $550$~K 
is due to the oxygen intake, the second one at $720$~K 
is due to the $\alpha\to\beta$ transition, and the last one at $860$~K is
due to the $\beta\to\gamma$ phase transition. The calculated
mass change during the phase transition at $550$~K 
is $0.65\%$, which is close to what is expected
from the oxidation of Bi$_{6}$V$_{3}$O$_{16}$ to Bi$_{4}$V$_{2}$O$_{11}$
($0.5\%$). 

Analysis of the V-V interaction path (see Fig.~\ref{fig:1 uniform chain structure})
reveals that two vanadium ions from crystallographic
site V$1$ (magnetic V$^{4+}$) are connected by two oxygen ions and
one vanadium from crystallographic site V$2$ (nonmagnetic V$^{5+}$).
The type of magnetic interaction between the V$^{4+}$ ions is hence 
extended to superexchange (V$^{4+}$-O$^{2-}$-V$^{5+}$-O$^{2-}$-V$^{4+}$).
The V-V distances are equal along the chains. V$^{4+}$ and V$^{5+}$
ions are in the pyramidal and tetrahedral environments of oxygens,
respectively (see Fig. \ref{fig:1 uniform chain structure}). 
The bond distances and bond angles along the V-V interaction
path for Bi$_{6}$V$_{3}$O$_{16}$ are listed in Tables \ref{tab3:Bond-lengths alfa}
and \ref{tab4:Bond-angles alfa}, respectively. From the structural point of view
the system resembles a $S=\frac{1}{2}$ uniform 1D Heisenberg chain.

\begin{table}
\begin{centering}
\begin{tabular}{|c|c|c|}
\hline 
Bonds  & Description  & Value (\AA)\tabularnewline
\hline 
\hline 
V$1$-O$5$  & intrachain  & 1.92\tabularnewline
\hline 
O$5$-V$2$  & intrachain  & $1.82$\tabularnewline
\hline 
V$2$-O$4$  & intrachain  & $1.73$\tabularnewline
\hline 
O$4$-V$1$  & intrachain  & $1.94$\tabularnewline
\hline 
\end{tabular}
\par\end{centering}

\centering{}\caption{\label{tab3:Bond-lengths alfa}Bond lengths between various vanadium-oxygen
linkages along the $a$ direction of the 1D chain in Bi$_{6}$V$_{3}$O$_{16}$ }
\end{table}

\begin{table}
\begin{centering}
\begin{tabular}{|c|c|c|}
\hline 
Angles  & Description  & Value (deg)\tabularnewline
\hline 
\hline 
V$1$-O$5$-V$2$  & intrachain  & 163.97\tabularnewline
\hline 
O$5$-V$2$-O$4$  & intrachain  & 101.07\tabularnewline
\hline 
V$2$-O$4$-V$1$  & intrachain  & 150.15\tabularnewline
\hline 
\end{tabular}
\par\end{centering}

\caption{\label{tab4:Bond-angles alfa}Bond angles between various vanadium-oxygen
linkages in Bi$_{6}$V$_{3}$O$_{16}$ }

\end{table}

Recent reports\cite{BiV2O5_Oct 2015_paper} on the crystallographic
data for the similar compounds Bi$_{4}$V$_{2}$O$_{10.2}$ and Bi$_{3.6}$V$_{2}$O$_{10}$
document that V$^{4+}$ ions occupy the V$1$ site, and V$^{5+}$
ions are placed on the V$2$ sites. Bi$_{6}$V$_{3}$O$_{16}$
is derived from the same Aurivillius family, having a similar composition and preparation route.
Here as well, V$^{4+}$ ions and V$^{5+}$ ions occupy the V$1$ and V$2$
sites, respectively. 

The field (up to 14~T) and temperature ($2~\lyxmathsym{\textendash}~300$~K) dependence
of the heat capacity and 
magnetization $M$ were measured using the 
heat capacity and vibrating-sample magnetometer options of a Quantum Design physical
property peasurement system, respectively.


The magic angle spinning nuclear magnetic resonance (MAS-NMR) measurements 
(spectra and nuclear spin-lattice relaxation times $T_{1}$) were carried out on vanadium 
nuclei ($^{51}$V gyromagnetic ratio $\lyxmathsym{\textgreek{g}}/2\lyxmathsym{\textgreek{p}}=11.1921$
MHz T$^{\lyxmathsym{\textminus}1}$ and nuclear spin $I=7/2$) in a fixed field of the $4.7$-T magnet
using an AVANCE-II Bruker spectrometer. The spinning speed of the $1.8$-mm o.d. rotor 
was varied between $20$ and $30$~kHz.\cite{MAS-NMR reference}
Typical pulse widths were varied from $4$ to $2$ $\mu$s. In echo
sequence $\lyxmathsym{\textgreek{p}}/2\lyxmathsym{\textminus\textgreek{t}\textminus\textgreek{p}}$, 
a rotation period between the excitation and refocusing
pulses was needed, \textgreek{t}. By measuring NMR using this technique, the nuclear dipole-dipole
interactions and chemical shift anisotropy are averaged
out, and the quadrupolar interaction is partially averaged out. Thus,
MAS-NMR gives finer details about the spectra. For $T_{1}$,
the speed of the rotor was $30$~kHz, and the measurements were undertaken
at the frequency of the isotropic shift at a given temperature. 
$T_{1}$ was measured by the saturation recovery method using a saturation
comb of fifty $\pi$/2 pulses followed by variable delay and an echo sequence
for recording the intensity. For the lowest temperatures, $T_{1}$ measurements were performed in 
static conditions to explore the low-temperature behavior down to $4$~K. The frequency 
shifts are given relative to the VOCl$_3$ reference.


\section{Results and Discussion}

\subsection{Magnetic susceptibility}

\begin{figure}
\begin{centering}
\includegraphics[scale=0.49]{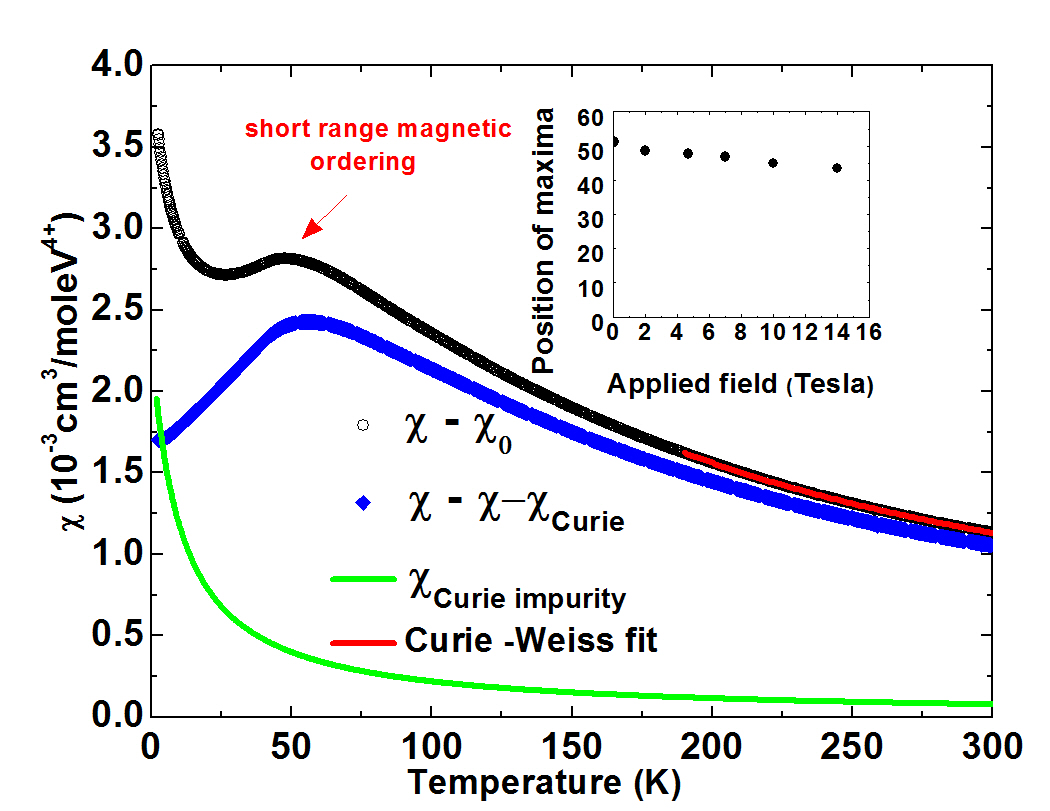}
\par\end{centering}

\caption{\label{fig:5_CW_Fit}The temperature dependence of the susceptibility
$\chi = M/H$ at $H = 4.7$~T for Bi$_{6}$V$_{3}$O$_{16}$.
Black open circles are $\chi~-~\chi_{0}$ data, the green line 
is the low-$T$ Curie-Weiss contribution and the blue solid diamonds show
$\chi~-~\chi_{0}~-~\chi_{Curie}$.
The region of short-range magnetic ordering is indicated by the red arrow.
The red line shows the high-$T$ Curie-Weiss fit in the temperature range of
$190~\lyxmathsym{\textendash}~300$~K. 
The inset shows the variation of the position of maximum in the magnetic
susceptibility curves with the change in applied field.}

\end{figure}
Magnetic susceptibility $\chi = M/H$ measurements were carried out in 0.05~-~14-T 
applied magnetic fields. With decreasing temperature, $\chi$ follows the 
Curie-Weiss law and shows a broad maximum around $50$~K, which indicates the presence of
short-range magnetic ordering in the system. A Curie-like upturn is
observed at lower temperatures, possibly arising from paramagnetic
impurities. From the Curie-Weiss fit, $\chi(T)=\chi_{0}+C/(T-\theta_{CW})$
in the $T$ range $190-300$ K (see Fig. \ref{fig:5_CW_Fit}), 
the $T$-independent  $\chi_{0}=8.6\times10^{-6}$
cm$^{3}$/(mole~V$^{4+}$), the Curie constant $C=0.4$ cm$^{3}$K/(mole~V$^{4+}$), and
the Curie-Weiss temperature\ \ $\theta_{CW}$=$-60$~K can be extracted. The negative
value of $\theta_{CW}$ indicates that the dominant exchange couplings
between V$^{4+}$ ions are AFM. We also measured the 
electron spin resonance spectrum of the powder ($X$ band, 
room temperature; not shown) and found typical g-values for a V$^{4+}$  ion  [($g_x, g_y, g_z$)= (1.93, 1.91, 1.8)] with tiny anisotropy.

From our $\chi$  measurements in Bi$_{6}$V$_{3}$O$_{16}$,
the value of the Curie constant ($C=0.135$ cm$^{3}$K/mole) is $36\%$
of the expected value ($C=0.375$ cm$^{3}$K/mole) for a full $S=1/2$
moment which indicates that only about $1/3$ of the vanadium ions
are magnetic, i.e., in the V$^{4+}$ oxidation state. Earlier 
reports~\cite{Ref.19_magnetic measurement-BiV2O5_2013,BiV2O5_Oct 2015_paper}
on similar systems (Bi$_{4}$V$_{2}$O$_{10.2}$ and Bi$_{3.6}$V$_{2}$O$_{10}$)
suggested that V$^{4+}$ and V$^{5+}$ ions prefer the V$1$ and V$2$
sites respectively, namely $90\%$ of the V$1$ site is
occupied by V$^{4+}$ and the rest is filled by the nonmagnetic V$^{5+}$,
meaning that V$1$ site is shared by the mixed
valences. From the bulk measurements,
we have not acquired any evidence of the existence of site sharing
in Bi$_{6}$V$_{3}$O$_{16}$. 

From the $\chi(T)$ results, the Van Vleck susceptibility
$\chi_{VV}=\chi_{0}-\chi_{core}=12.4\times10^{-5}$cm$^{3}$/mole,
where $\chi_{core}$ is the core diamagnetic susceptibility, was 
calculated with a value of $-11.52\times10^{-5}$cm$^{3}$/mole f. u. (here we 
consider the formula unit to be Bi$_{2}$VO$_{5.33}$). The low-temperature
upturn in $\chi(T)$ below $20$ K is attributed to the orphan spins
and other extrinsic magnetic impurities.~\cite{Lebernegg2017}

To extract the exact magnetic susceptibility, the temperature-independent
susceptibility (Van Vleck plus the core diamagnetic susceptibility)
and the Curie contribution originating from the extrinsic paramagnetic
impurities and/or orphan spins were subtracted from the experimentally
obtained magnetic susceptibility data (see Fig. \ref{fig:5_CW_Fit}). 
The low-temperature Curie-Weiss fit gives $C~=~0.022$~cm$^{3}$K/mole, 
$\theta_{CW}$~=~$10.3$~K. From the value of $C$ we find that this contribution is about 6\% 
of an ideal $S=\frac{1}{2}$ system.  After subtracting this, 
we find that the magnetic susceptibility saturates to a fixed value as $T$ tends 
towards zero, which is expected for a gapless $S=\frac{1}{2}$ uniform chain.
However, to model the susceptibility with the uniform $S=\frac{1}{2}$ Heisenberg chain, 
we rely on the MAS-NMR data below, where the spin susceptibility is 
manifested in a more pristine manner.

\begin{figure}

\begin{centering}
\includegraphics[scale=0.32]{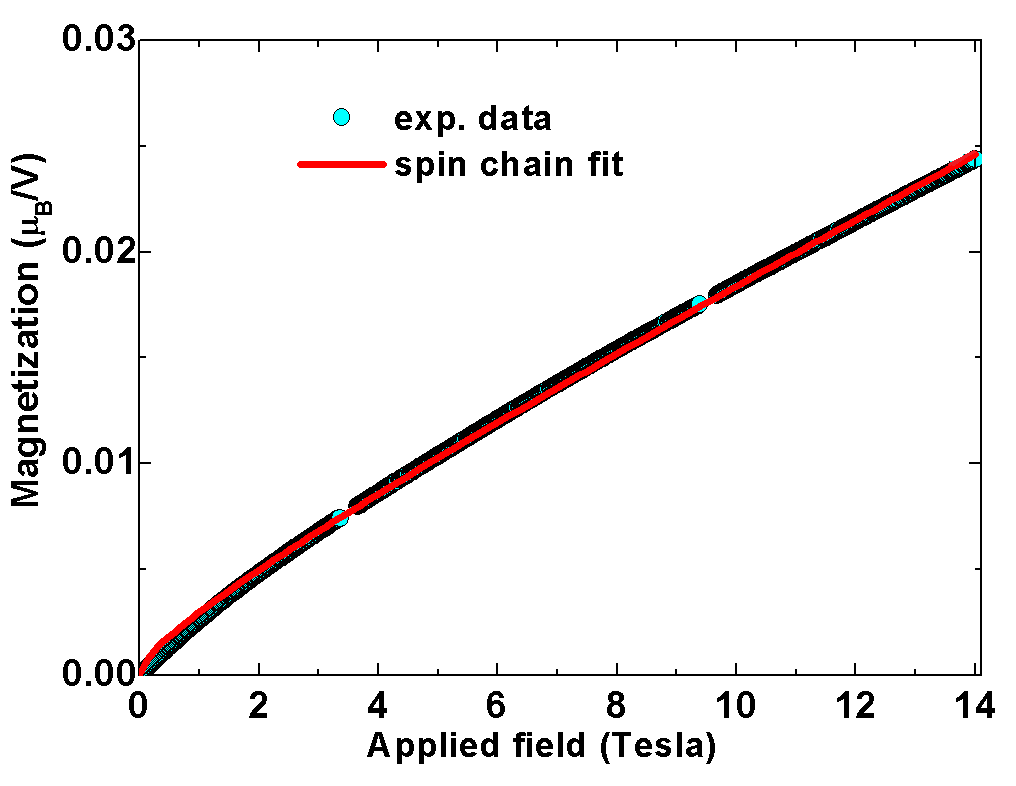}
\par\end{centering}

\caption{\label{fig:6 Mvs H}The experimental magnetization data of Bi$_{6}$V$_{3}$O$_{16}$
vs applied magnetic field at 1.8~K. }

\end{figure}

We have not observed any signature of hysteresis in the $M$ vs $H$ measurements.
The broad maximum observed in our system greatly resembles
what was observed in similar Bi-V-O complexes reported previously.\cite{BiV2O5_Oct 2015_paper,Ref.19_magnetic measurement-BiV2O5_2013}
With the increase in the applied field, we observe that the broad maximum
shifts towards lower temperatures (see the inset of Fig.~\ref{fig:5_CW_Fit}).

In the applied field $H$ dependence of magnetization $M$ 
(see Fig.~\ref{fig:6 Mvs H}), we have not found any anomaly
or steps indicating the presence of any gap, and the 
data agree well with the phenomenological expression 
$M^{\text{chain}}(H) = \alpha H + \beta\sqrt{H}$,
with  $\alpha$= 1.3$\times10^{-7}$, and $\beta$ = 1.65$\times10^{-5}$.

\subsection{Heat capacity}

\begin{figure}
\begin{centering}
\includegraphics[scale=0.59]{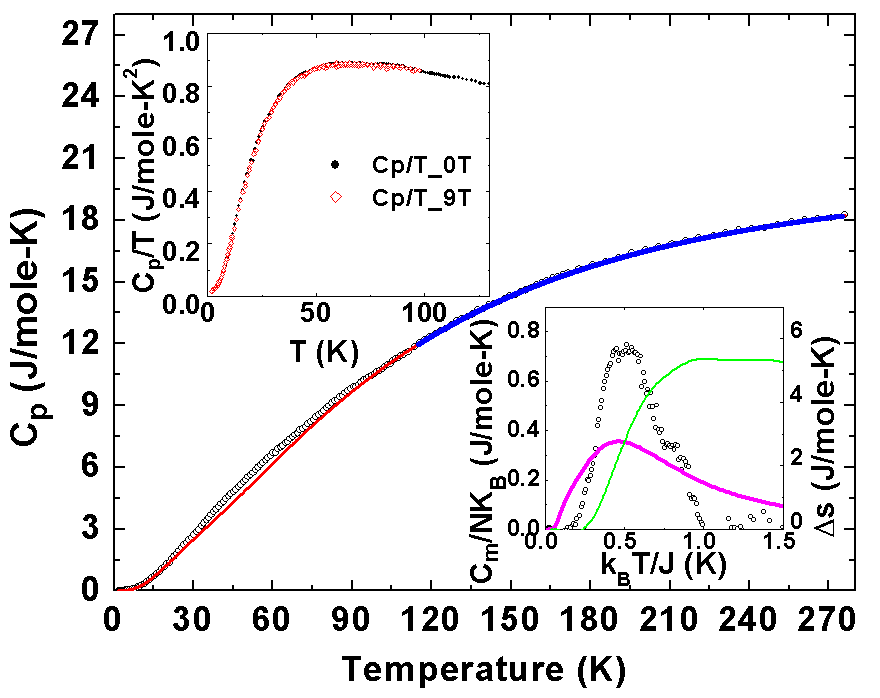}
\par\end{centering}

\caption{\label{fig:7_heat capacity}The temperature dependence of specific
heat of Bi$_{6}$V$_{3}$O$_{16}$ 
in zero magnetic field; the
blue points are the fit described in the text, and the red line is its
extrapolation. The top inset displays the $C_{p}/T$ vs T plot
at $0$ T (black solid circles) and $9$ T (red open diamonds) magnetic
fields, respectively. 
The bottom inset shows the magnetic contribution to the heat capacity (black
open circles). The magenta line is the magnetic heat capacity contribution 
for the 1D uniform Heisenberg chain; the green data points (right axis) show the change
in entropy $\Delta S$ with $T$. }
\end{figure}

In the temperature dependence of heat capacity $C_{p}$, we did not detect 
any long range magnetic ordering (Fig.\ref{fig:7_heat capacity}).
In the plot of $C_{p}/T$ vs. $T$, the broad maximum at around $55$ K is observed,
which does not shift under the application of external
magnetic field up to $9$ T. The magnetic specific heat $C_{m}$ was extracted 
by subtracting the lattice contribution using a combination
of Debye and Einstein heat capacities, $C_{Debye}$ and $C_{Einstein}$,
respectively:

$C_{Debye}=\ensuremath{C_{d}\times9nR}\ensuremath{(T/\theta_{\text{d}})^{\mbox{\text{3}}}}\ensuremath{\int_{\text{0}}^{\text{\ensuremath{\theta\ensuremath{_{\text{d}}}/T}}}[x^{\text{4}}e^{\text{\ensuremath{x}}}/(e^{\text{\ensuremath{x}}}-1)^{\text{2}}}\ensuremath{]dx}$,

$C_{Einstein}=\ensuremath{3nR[\sum C{}_{e_{m}}\frac{x_{E_{m}}^{2}e^{x_{E_{m}}}}{(e^{x_{E_{m}}}-1)^{2}}]}$,
$x=\frac{h\omega_{E}}{k_{B}T}$.

In the above formula, $n$ is the number of atoms in the primitive
cell, $k_{\text{B}}$ is the Boltzmann constant,  $\theta_{\text{d}}$
is the relevant Debye temperature, and $m$ is an index for an optical
mode of vibration. In the Debye-Einstein model the total number of
modes of vibration (acoustic plus optical) is equal to the total number
of atoms in the formula unit. In this model we considered the
ratio of the relative weights of acoustic modes and the sum of the different
optical modes to be $1:n-1$. 

We used a single Debye and multiple (three) Einstein functions
with the coefficient $C_{d}$ for the relative weight of the acoustic modes of 
vibration and coefficients
$C_{e_{1}}$, $C_{e_{2}}$, and $C_{e_{3}}$ for the relative weights
of the optical modes of vibration. 
The experimental data of the system were fitted by excluding the low-temperature
region of $2-115$ K assuming that most of the magnetic part of the
heat capacity is confined within this temperature range.
The fit of our experimental data
to such a combination of Debye and Einstein heat capacities
yields a Debye temperature of $96$~K and Einstein temperatures of
$130$, $295$, and $584$~K with relative weights of $C_{d}$
:$C_{e_{1}}$:$C_{e_{2}}$:$C_{e_{3}}$ = $13:14:48:25$.

The electronic contribution to the
total heat capacity was neglected since the compound possesses an 
insulating ground state. Upon subtracting
the lattice heat capacity with the above parameters, we obtain the
magnetic contribution to the heat capacity $C_{m}(T)$, which, accordingly, 
shows a broad maximum around $50$ K. The entropy
change $\lyxmathsym{\textgreek{D}}$$S$ was calculated by integrating
the $C_{m}/T$ data (see bottom inset in Fig.~\ref{fig:7_heat capacity}). The entropy
change is about $5.36$~J K$^{\mbox{\textminus}1}$ (calculated for
1 mole), which is close to the expected value of a $S=\frac{1}{2}$ system
($R$ ln 2 = $5.73$~J/mole~K). Although the observance of the broad maximum
in the $C_{m}$ vs $T$ data indicates the 1D magnetic interaction
in the system, in the temperature regime 
below $30$~K, the magnetic heat capacity is less than 1 \%
of the lattice contribution, which makes the analysis of the magnetic
contribution in this regime highly dependent on the model. 

We have also compared $C_{m}$ with the 1D Heisenberg chain 
model~\cite{Alternating chain expression paper_Johnston} 
(see the magenta line in bottom inset in Fig.~\ref{fig:7_heat capacity}). 
The mismatch between our calculated $C_{m}$ and the fit is not surprising 
as we know that an accurate estimation of $C_{m}$ is nearly impossible 
at this temperature range as the phonon contribution (lattice part) consists 
of nearly 99\% of the heat capacity around the peak of $C_{m}$ at such a relatively high temperature.

The main findings from our heat capacity results are that we did not observe 
any magnetic long-range ordering in our system and $C_{m}$ shows the same 
trend observed in  $\chi$  vs $T$ data. Both these results support the 
low-dimensional magnetic behavior in Bi$_{6}$V$_{3}$O$_{16}$.

\subsection{NMR}

\subsubsection{Room-temperature magic angle spinning NMR}

\begin{figure}
\begin{raggedright}
\includegraphics[scale=0.52]{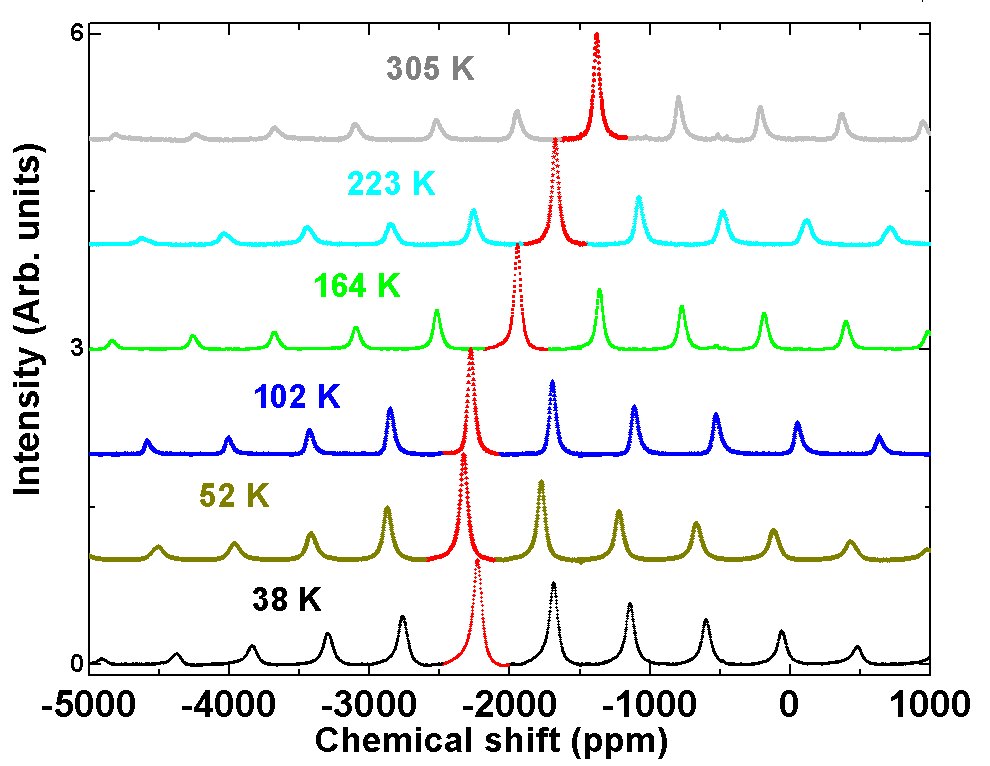}
\par\end{raggedright}

\caption{\label{fig:8_MASNMR_spectra}$^{51}$V MAS-NMR spectra of Bi$_{6}$V$_{3}$O$_{16}$ measured at different temperatures in 4.7-T fixed magnetic field. The main line at the isotropic value of the magnetic shift is highlighted in red, and the rest of the peaks are spinning sidebands at multiples of the spinning frequency (30 kHz) apart from the main line.}
\end{figure}

\begin{figure}
\begin{centering}
\includegraphics[scale=0.8]{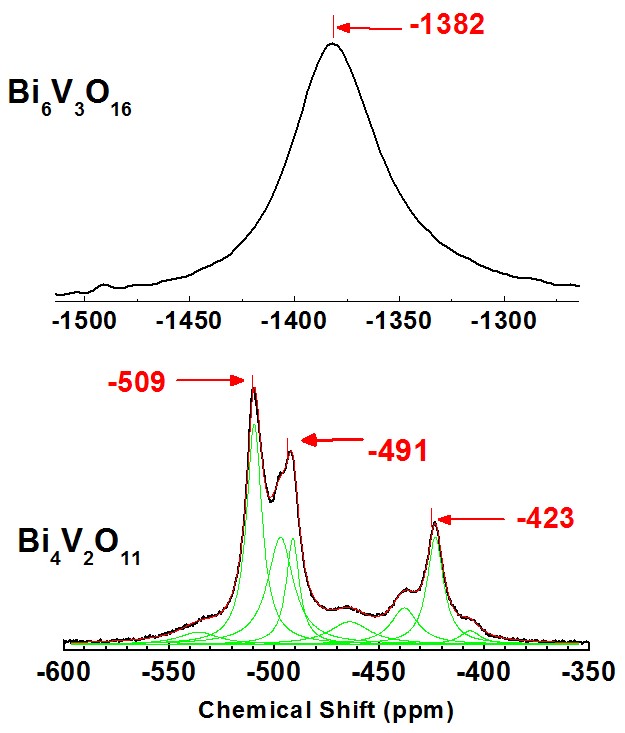}
\par\end{centering}

\caption{\label{fig:9 Bi4V2O11 NMR} The top spectrum is the zoomed part of
the central peak of the room-temperature ($305$ K) $^{51}$V NMR
spectrum of Bi$_{6}$V$_{3}$O$_{16}$. 
The bottom spectrum is the $^{51}$V
MAS-NMR spectrum of Bi$_{4}$V$_{2}$O$_{11}$. The black line is the
experimental spectrum, which can be well fitted by several Lorentzian
lines, as given by green lines; the red line is the sum of the components.
The spectra were recorded on an AVANCE-III-800 spectrometer at $^{51}$V
resonance frequency of $210.5$ MHz, using a home-built MAS probe for
$1.3$-mm rotors, at $50.1$-kHz sample spinning speed.}

\end{figure}

NMR is a powerful local probe to extract the static and dynamic properties
of a spin system and has been extensively used on vanadia systems.~\cite{Volkova}
Fortunately, the room-temperature $^{51}$V MAS-NMR spectrum for Bi$_{6}$V$_{3}$O$_{16}$
is known, consisting of a single line shifted to -1447 ppm at room temperature (with sample spinning
speeds up to 17~kHz).~\cite{Delmaire} In our MAS spectra, we 
observed also only one $^{51}$V line at -1382 ppm (spinning
at 30 kHz; see the uppermost spectrum in Fig.~\ref{fig:8_MASNMR_spectra}
and the top spectrum of Fig.~\ref{fig:9 Bi4V2O11 NMR}) confirming
that the $^{51}$V NMR signal originates entirely and from only one of the 
two available vanadium sites in this system.~\cite{comment1}
As the spectral resolution is much better in the $^{51}$V MAS-NMR
compared to any bulk measurement or even to the static NMR data, having
a single strongly shifted sharp NMR line validates the structure of
Bi$_{6}$V$_{3}$O$_{16}$ here very strongly.

We have also performed room-temperature MAS-NMR measurements on Bi$_{4}$V$_{2}$O$_{11}$,
the nonmagnetic parent compound of the BiMeVOX family, and compared the
$^{51}$V NMR signals of these two compounds under the same experimental
conditions (Fig.~\ref{fig:9 Bi4V2O11 NMR}). For Bi$_{4}$V$_{2}$O$_{11}$, the $^{51}$V MAS-NMR line
positions are also documented~\cite{Hardcastle,KimGray}, and
our results agree well with the literature. MAS-NMR results published
by Delmaire \textit{et~al.}\cite{Delmaire} showed the detection of four
(three major) structurally different V$^{5+}$ environments.
The MAS-NMR results of Kim and Grey showed the detection of three
different vanadiums.\cite{KimGray} 
Based on the crystal structure analysis of Mairesse\textit{ et~al.}\cite{structural analysis of Bi4V2O11}
and NMR studies of Kim and Grey \cite{KimGray} we can assign the 
peak at -423~ppm to the tetrahedral V$^{5+}$ (the V1 site according to the description of 
Ref.~\onlinecite{structural analysis of Bi4V2O11}), the peak at -509~ppm to V$^{5+}$ 
in the trigonal bipyramidal environment (V2), and peaks at -491 and -497~ppm to
the two different fivefold V$^{5+}$ environments. These last two are V3a and V3b according to the report by Mairesse \textit{ et~al.};\cite{structural analysis of Bi4V2O11} however, we cannot differentiate  
which one is V3a and which is V3b in our spectrum. The remaining lines with lower intensities are possibly due to the V$^{5+}$ ions close to the ends of chains and from the $6a_{m}$ superstructure which was detected at 
low level in the XRD data of Ref.~\onlinecite{structural analysis of Bi4V2O11}.
The chemical shift, the width, and relative intensity of these components
are given in Table~\ref{tab:4Bi4V2O11 components}.

The determined aspects further validate that in the
case of Bi$_{6}$V$_{3}$O$_{16}$, the possibility of V$^{4+}$
ions occupying two different sites is clearly ruled out, as this would
lead to the creation of different environments of V$^{5+}$ ions and,
consequently, different NMR lines which should get
detected in MAS-NMR experiments. However, due to a strong,
large hyperfine field on V$^{4+}$ ions, the NMR signal originating
from the magnetic vanadium ions could not be detected at elevated
temperatures. A similar scenario was observed in many
other systems such as Cs$_{2}$CuCl$_{4}$,\cite{Cs2CuCl4} BaCuSi$_{2}$O$_{6}$,
\cite{Jaime04,Kramer07,Kramer13} SrCu$_{2}$(BO$_{3}$)$_{2}$,\cite{Kageyama99,Kodama02}
BaV$_{3}$O$_{8}$,\cite{BaV3O8} and Li$_{2}$ZnV$_{3}$O$_{8}$.\cite{Li2ZnV3O8}
The example of BaV$_{3}$O$_{8}$ is most relevant in this context
because BaV$_{3}$O$_{8}$ is also a 1D chain system where the
signal from the magnetic V$^{4+}$ ions was not detected, while the
signal from the nonmagnetic V$^{5+}$ was observed.\cite{BaV3O8}

\begin{table}

\begin{centering}
\begin{tabular}{|c|c|c|c|}
\hline 
 & Frequency& FWHM&Relative\\&shift (ppm) &(ppm)&intensity (\%)\tabularnewline
\hline 
\hline 
Peak 1 & $-423$ & $10$ & $15$\tabularnewline
\hline 
Peak 2 & $-509$ & $9$ & $29$\tabularnewline
\hline 
Peak 3 & $-491$ & $8$ & $11$\tabularnewline
\hline 
Peak 4 & $-438$ & $16$ & $8$\tabularnewline
\hline 
Peak 5 & $-407$ & $12$ & $2$\tabularnewline
\hline 
Peak 6 & $-464$ & $23$ & $8$\tabularnewline
\hline 
Peak 7 & $-536$ & $23$ & $4$\tabularnewline
\hline 
Peak 8 & $-497$ & $15$ & $23$\tabularnewline
\hline 
\end{tabular}
\par\end{centering}

\caption{\label{tab:4Bi4V2O11 components}The chemical shift, the width, and
relative intensity of different vanadium lines in Bi$_{4}$V$_{2}$O$_{11}$spectra. }
\end{table}

\subsubsection{Low-temperature, cryoMAS NMR}

The temperature dependence of $^{51}$V spectra of Bi$_{6}$V$_{3}$O$_{16}$
measured using cryogenic MAS (cryoMAS) technique~\cite{MAS-NMR reference} is shown in 
Fig.~\ref{fig:8_MASNMR_spectra}.
Here, we are limited to remain above $T=$ $20$~K to maintain the
fast sample spinning, but the obtained values of the isotropic 
Knight shift $K$ up to room temperature
are very accurately determined, and they follow the same trend
observed in $\chi(T)$. In the temperature
dependence of $K$, a broad maximum at around $50$ K 
is observed, similar to the $\chi(T)$ data, indicating low-dimensional,
short-range magnetic ordering.

\begin{figure}
\begin{centering}
\includegraphics[scale=0.38]{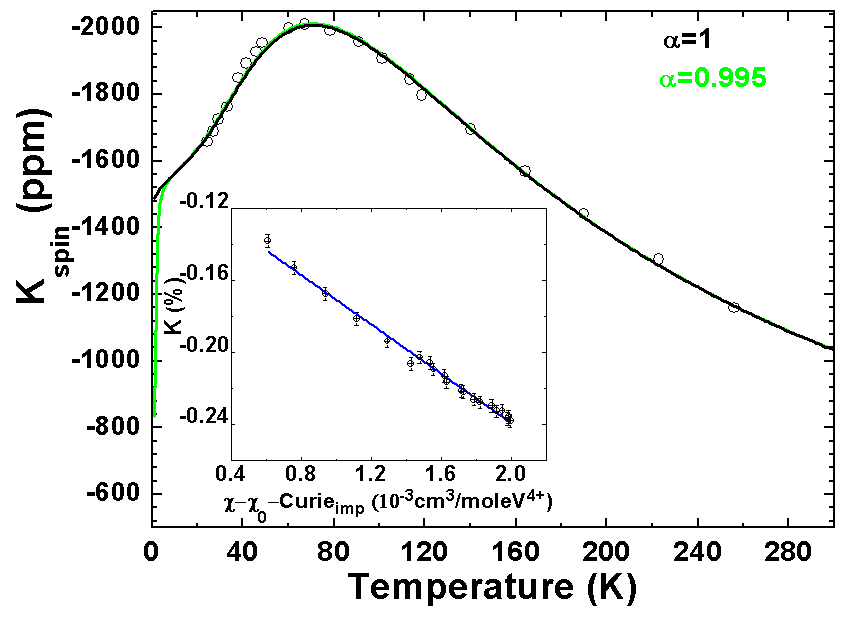}
\par\end{centering}

\caption{\label{fig:10_MASNMR_K_T}Temperature dependence of the $^{51}$V NMR
shift $K$ of Bi$_{6}$V$_{3}$O$_{16}$
(shown by black
circles) measured using the cryoMAS technique. The black and green lines are the fittings with
the susceptibility model using Eq.(~\ref{eq:2}) for a $S=\frac{1}{2}$ uniform
chain ($\alpha=1$) and for an alternating chain ($\alpha=0.995$), respectively.
The inset shows the $^{51}$V MAS-NMR shift  $K$ of Bi$_{6}$V$_{3}$O$_{16}$ 
vs $\chi-\chi_{0}-\chi_{Curie}$, where both $\chi(T)$ and $K$ are measured at
$4.7$~T, with temperature being an implicit parameter. The solid line
is the linear fit.}
\end{figure}

As $K(T)$ is a direct measure of spin susceptibility, the following equation can 
be written:
\begin{equation}
K(T)=K_{0}+\frac{A_{hf}}{N_{A}\mu_{B}}\chi_{spin}(T)\label{eq:2}~,
\end{equation}
where $K_{0}$ is the temperature-independent chemical shift, $A_{hf}$
is the hyperfine coupling constant, and $N_{A}$ is Avogadro's
number. As long as $A_{hf}$ is constant, $K(T)$ should follow $\chi_{spin}(T)$.
We estimated the exchange couplings by fitting the $K(T)$ data with 
Eq.(~\ref{eq:2}). Here, $\chi_{spin}$ is the expression for the spin susceptibility
of the $S=\frac{1}{2}$ chain model given by 
Johnston \textit{et~al.}\cite{Alternating chain expression paper_Johnston}\textit{
} which is valid in the whole temperature range of our experiment from
$2$ to $300$~K and also in the whole limit of $0\leqslant\alpha\leqslant1$.
The $K(T)$ data for Bi$_{6}$V$_{3}$O$_{16}$ agree well with the
$S=\frac{1}{2}$ chain model with $K_{0}\simeq-370$~ppm, $A_{hf}=5.64$ KOe/$\mu_{B}$
with an exchange coupling $J_{1}/k_{B}=113(5)$ K, and the alternation
ratio $\alpha=1$ (uniform chain) and $\alpha=0.995$ (alternating
chain; Fig.~\ref{fig:10_MASNMR_K_T}). 
For the alternation ratio of $\alpha=0.95$,
the zero-field spin gap between the singlet and triplet states according
to the $S=\frac{1}{2}$ alternating chain model is $\Delta/k_{B}\simeq9.52$
K according to Johnston \textit{et~al.}\cite{Alternating chain expression paper_Johnston}\textit{
} and $10.35$ K according to Barnes \textit{et~al.},\cite{spin gap in alternating spin chain}\textit{
} depending on the method of approximation. Our $M$ vs $H$ results up to $14$~T 
(see Fig.~\ref{fig:6 Mvs H}) did not show any signature of closing
of the spin gap near the respective magnetic fields, which prompts us to consider
that uniform chain is the correct model indeed. 

It is insightful to compare our results with the recently investigated uniform chain compound
(NO)[Cu(NO$_3$)$_3$]  with intrachain coupling $J~=~142(3)$\,K and 
long-range magnetic order taking place only at $T_N~=~0.58(1)$\,K, resulting in a ratio of 
suppression represented by $f~=~|J|/T_N~=~245(10)$.~\cite{NOCu(NO3)3-1,NOCu(NO3)3-2,NOCu(NO3)3-3}
Even more recently, a study on  Cs$_4$CuSb$_2$Cl$_{12}$ reported $J~=~186(2)$\,K and 
a superconductorlike phase transition taking place only at $T_{sp}~=~0.70(1)$\,K, resulting in a ratio of 
suppression represented by $f~=~|J|/T_{sp}~=~270(7)$.~\cite{Tran2019} 
KCuF$_3$ has a relatively large interchain coupling ($J'/J\approx0.01$), yielding 
$f~=~390$~K$/39$~K$~=~10$;\cite{Lake2005}\   Sr$_2$CuO$_3$, with a tiny interchain 
interaction ($J'/J\approx10^{-5}$), gives $f~=~2200$~K$/5$~K$~=~440$.\cite{SrCu2O3}
For Bi$_{6}$V$_{3}$O$_{16}$ the lowest estimate would be $f~=~108$~K$/2$~K$~\approx~55$, 
suggesting that the interchain exchange interactions here are very weak and/or frustrated.

The $K$ vs $\chi(T)$ plot is shown in the inset of Fig.~\ref{fig:10_MASNMR_K_T}, 
where $K$ is measured shift in percent and $\chi-\chi_{0}-\chi_{Curie}$
is magnetic susceptibility without the $T$-independent and Curie impurity contributions. 
The magnetic susceptibility was measured in the same magnetic field of $4.7$~T in 
which the NMR measurements were performed.

\subsubsection{Spin lattice relaxation rate $1/T_{1}$}

\begin{figure}
\begin{centering}
\includegraphics[scale=0.54]{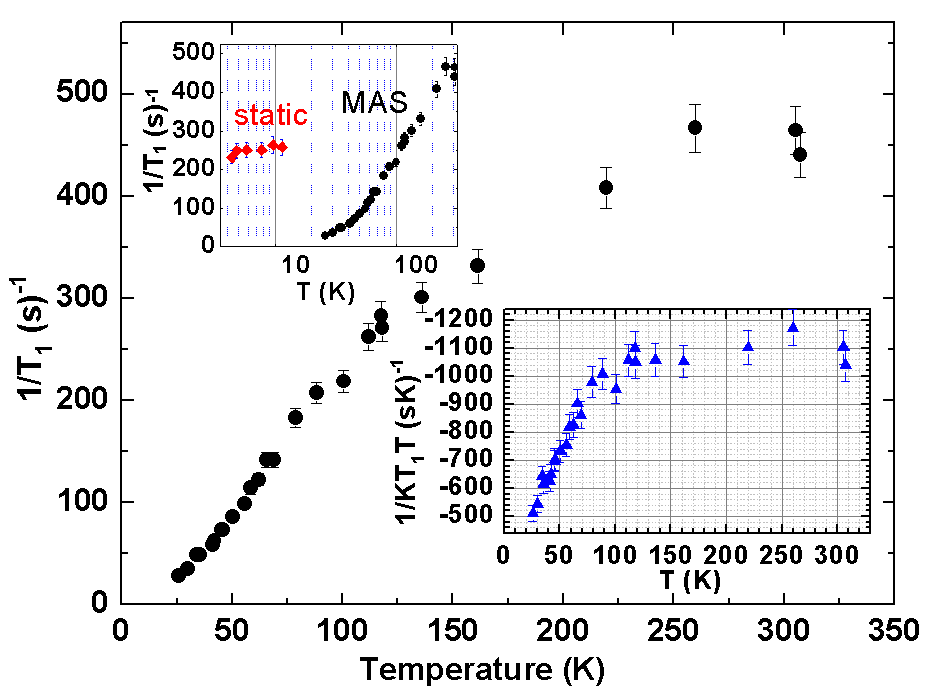}
\par\end{centering}

\caption{\label{fig:11_T1}Temperature dependence of the spin-lattice relaxation
rate ($1/T_{1}$) of Bi$_{6}$V$_{3}$O$_{16}$. \ 
In the top inset the same plot is shown with temperature in the log scale and low-$T$
data measured on a static sample (red diamonds). The bottom
inset shows the plot of $1/KT_{1}$T vs $T$.}
\end{figure}

To study microscopic properties of 1D Heisenberg antiferromagnets (HAF), it is necessary
to measure the temperature dependence of the spin lattice relaxation rate
$1/T_{1}$, which gives information about the imaginary part of the
dynamic susceptibility $\chi(q,\omega)$. As vanadium is a $I=7/2$
nucleus and to avoid further broadening due to dipole-dipole interaction
we studied the temperature dependence of $1/T_{1}$ in the rotating
conditions. The temperature dependence of $^{51}$V $1/T_{1}$ is presented
in Fig.~\ref{fig:11_T1}. In the whole temperature range from $300$
down to $20$~K, the recovery of nuclear magnetization is single exponential.
We have not observed any indication of divergence of the relaxation
rate, revealing the absence of any magnetic ordering. Also, no
sign of activated behavior was observed, which proves that
down to $20$~K no dimerization takes place. Until about $100$ K,
$1/T_{1}$ drops linearly with temperature. However, below 100~K deviations
from the linear behavior are observed. 
Additionally, $1/KT_{1}T$ is temperature independent at $100$~K~$\leq~T~\leq~300$~K, 
and it shows linear behavior with $T$ below $100$~K.

We did not observe any signatures of magnetic ordering, and also no
features of spin gap are observed in the temperature dependence of $1/T_{1}$.
Generally, $1/T_{1}$ depends on both uniform $(q=0)$ and staggered
spin fluctuations $(q=\pm\pi/a)$. The uniform component leads to
$1/T_{1\:}\sim\: T$, while the staggered component gives $1/T_{1}$$=$
const.\cite{spin fluctuations-Sachdev} The deviation from the linear behavior
of $1/T_{1}$ below $100$ K presumably indicates that the temperature-independent
part is coming into play which was otherwise absent in the temperature
region above $100$ K. This fact is also reflected in the $1/KT_{1}T$
vs $T$ plot as $1/KT_{1}T$ is expected to be constant when the ($q=0)$
contribution dominates. In our  $1/KT_{1}$T plot, we observe a clear drop 
from the high-$T$ constant value for temperatures $\leq$100 K. 

To observe the temperature dependence of $1/T_{1}$ at lower temperatures, we 
stopped spinning and  performed NMR measurements on the broad line 
in static conditions; $1/T_{1}$ below $15$~K is essentially $T$ independent.
The absolute value of the relaxation rate at low $T$ is much larger than in 
cryoMAS-NMR, which seems to indicate that
in the static $T_{1}$ measurements we have started to detect the magnetic 
V$^{4+}$ at low temperatures with a very short relaxation 
(see the top inset in Fig.~\ref{fig:11_T1}). Note that $1/T_{1}$ 
becoming constant at low temperatures is expected for $S=\frac{1}{2}$ 1D 
HAF systems.~\cite{Barzykin} Similar spin-lattice
relaxation behavior has been observed in the uniform $S=\frac{1}{2}$ 1D
chain Ba$_{2}$Cu(PO$_{4}$)$_{2}$.\cite{Ba2Cu(PO4)2-NMR-T1}



\section{Conclusion and Outlook}

We have reported bulk thermodynamic and local NMR studies of the $S=\frac{1}{2}$ V-based
compound Bi$_{6}$V$_{3}$O$_{16}$. All of the measurements
confirm the presence of low dimensionality in this material. 
Upon subtracting the low-temperature Curie-Weiss contribution, the magnetic 
spin susceptibility agrees well with the $S=\frac{1}{2}$ uniform Heisenberg chain model.
The magnetic heat
capacity also confirms the existence of low-dimensional magnetism
in the system, even though the lattice part has a dominant
contribution to the total heat capacity, and approximation by any
model is not decisive. In the MAS-NMR experiments on Bi$_{6}$V$_{3}$O$_{16}$,
we observed a single sharp line which confirms that there is no site
sharing between the V$^{4+}$ and V$^{5+}$ ions in this compound.
The spin susceptibility calculated from the MAS-NMR experiments agrees well 
with the uniform $S=\frac{1}{2}$ chain model with the dominant exchange coupling 
of $J = 113(5)$~K, while the temperature variation of Knight shift 
agrees well with 
the findings from $\chi$ vs $T$ measurements.
Our experimental results from cryoMAS-NMR measurements concur 
with the $S=\frac{1}{2}$ uniform chain model up to the available temperature range. 
No sign of magnetic ordering
or any feature of spin gap has been observed in the temperature dependence
of $1/T_{1}$. 
Bi$_{6}$V$_{3}$O$_{16}$ is one of the very few V-based systems in the category 
of uniform spin chains where no long-range magnetic ordering or any singlet
formation were observed above $2$~K, while $J$ is of the order of $100$~K.
An ideal $S=\frac{1}{2}$ spin chain cannot exist in any real material because even an infinitesimal 
interchain coupling would give rise to long-range magnetic order at suppressed, but 
finite, temperatures. 

Future work involving 
local probe experiments, e.g, static NMR experiments down to millikelvin
temperatures, neutron diffraction, muon spin resonance, etc., 
is needed to acquire further knowledge about the possible ordering temperature and 
the nature of the ordered magnetic structure
of Bi$_{6}$V$_{3}$O$_{16}$, for which a high-quality single crystal is needed.

\section{acknowledgement}

We thank M. B. Ghag and P. N. Mundye for the help and support during the sample
preparation at TIFR, A. A. Tsirlin and K. Kundu for insightful discussions, K. T\~{o}nsuaadu for
help with TG/DTA experiments and analysis,
and T. Tuherm for his help during our cryoMAS measurements
at NICPB. We acknowledge the Department of Science and Technology,
government of India; European Regional Development Fund (TK134); Estonian research council (PRG4, IUT23-7, MOBJD295, MOBJD449) for financial support.

\end{document}